\documentclass[twocolumn]{svjour3}         % twocolumn
\smartqed  % flush right qed marks, e.g. at end of proof
\usepackage[dvips]{graphicx}
\journalname{Granular Matter}

\begin{document}

\title{Motion of grains in a vibrated U-tube}

\author{J.R. Darias\and I. S\'{a}nchez \and G. Guti\'{e}rrez}

\institute{J.R. Darias, G.Guti\'{e}rrez \at
Departamento de F\'{\i}sica, Universidad Sim\'{o}n Bol\'{\i}var, Apartado Postal 89000, Caracas 1080-A, Venezuela.\\
\email{jrdarias@usb.ve}\\
\email{gustav@usb.ve}\\
Tel. +58-212-9063541\\
Fax. +58-212-9063601\\
%\emph{Present address:} of F. Author  %  if needed
\and
I. S\'{a}nchez \at Centro de F\'{\i}sica, Instituto Venezolano de Investigaciones Cient\'{\i}ficas, Apartado Postal 21827, Caracas 1020-A, Venezuela.\\
\email{ijsanche@ivic.ve}\\
Tel. +58-212-5041534}
\date{Received: date / Accepted: date}

\maketitle

\begin{abstract}

We investigate experimentally the behavior of the rate of growth of a column of grains, in a partially filled vertically shaken U-tube. For the set of frequencies used we identify three qualitatively different behaviors for the growth rate $\gamma$ as a function of the dimensionless acceleration  $\Gamma$: 1) an interval of zero growth for low $\Gamma$ with a smooth change to nonzero growth, analogous to a continuous phase transition; 2) a sigmoidal region for $\gamma$ at intermediate values of the dimensionless acceleration $\Gamma$; and 3) an abrupt change from high values of $\gamma$ to zero growth at high values of $\Gamma$, similar to a first order phase transition. We obtain that our data is well described by a simple differential equation for the change of the growth rate with the dimensionless acceleration of the vertical vibrations.

%\PACS{PACS code1 \and PACS code2 \and more}
%\subclass{MSC code1 \and MSC code2 \and more}
\end{abstract}

\section{Introduction}

The collective rise of grains in one branch of a vertically vibrated U-tube and some related instabilities like heaping and granular transport in vibrated granular deep beds in partitioned containers has been the subject of research for decades \cite{Gutman,Evesque,Rajchenbach,King,Ivan,Akiyama,Chen,Maeno}. In this work we concentrate in the experimental characterization of the rate of growth of the rising column of grains in a partially filled U-tube for different frequencies. The mechanism responsible for this interesting behavior is still an open issue.

In a previous work \cite{Ivan} the behavior of a vertically vibrated granular system in a partially filled U-tube at low frequency was studied. A collective granular transport upward through one of the branches of the tube was observed for small grains. The experimental results were compared with a model based on the idea of cyclic fluidization \cite{Gutierrez}. In that model it was assumed that for low frequencies and sufficiently high amplitudes of oscillations the granular bed fluidizes cyclically in such a way that an effective upward acceleration acts on the grains while these are in a fluidized state. Consequently, an instability appears  and the free surface of the bed on either branch of the U-tube rises while the other goes down. Therefore, one of the branches of the tube is rapidly filled and simultaneously the other empties. That model captures some relevant aspects of the observed behavior for low frequencies, as for example the exponential growth of the granular column and the monotonic increase of the growth rate with the dimensionless acceleration of the vibration; however, we have found some important quantitative discrepancies. The above model for the measured growth rate $\gamma$ versus the maximum dimensionless acceleration $\Gamma$ of the container provides only a crude approximation for low frequencies. As the frequency is increased above $10\,Hz$, it becomes clear that the model is insufficient to describe this phenomenon.

In this paper we extend the work in reference \cite{Ivan} by exploring the growth rate of the height difference between the two free surfaces, for a wider range of frequencies. We obtain a differential equation for the change of the measured growth rates with the dimensionless acceleration of the vibrations.

\section{Experimental setup}

For the vibrational system we used a function generator (GW Instek SFG-2110) linked to an audio amplifier (Crown XLS 202) feeding two coupled 1000 Watts loudspeakers. For the granular material we used glass spheres with diameter between $250$ and $300\,\mu m$ and bulk static density $\rho_g = 1440\,kg/m^3$. The top of the container was made permeable to air but not to grains (using a $125\,\mu m$ nylon mesh). The container is a U-shaped tube $200\,mm$ tall, with square internal cross section of $400\,mm^2$. The total mass of grains used was $124\,g$, chosen to completely fill one of the U-tube branches.

\begin{figure}[!ht]
\begin{center}
\includegraphics{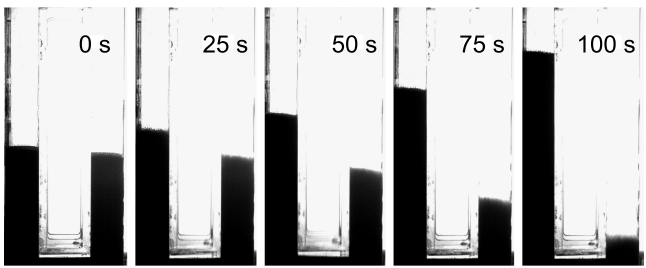}
\caption{Snapshot sequence of the partially filled U-tube in a typical experiment. The grains migrate to the branch of the tube with the initially taller granular column. The images shown correspond to an experiment for which the U-tube was shaken at a frequency of $20\,Hz$ and $\Gamma = 4.5$. The increase of the growth velocity with time can be readily appreciated.}\label{fig1}
\end{center}
\end{figure}

Images of the experiments were recorded using a digital camera (Pixelink PL-B741F). Filming against a back light (see Fig. \ref{fig1}), allowed us to analyze the images using a commercial software to obtain the difference in height $\Delta h$ between the two free surfaces as a function of time.

\begin{figure}[!ht]
\begin{center}
\includegraphics{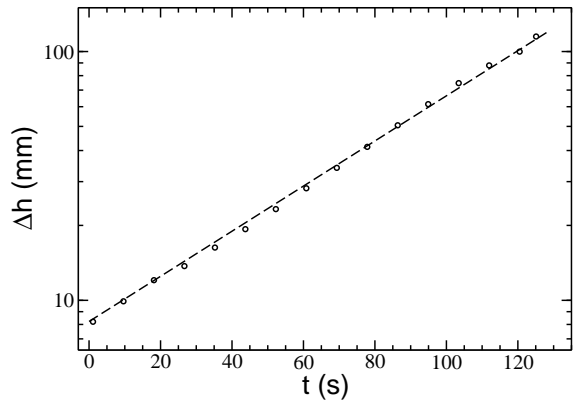}
\caption{Typical time dependence of the height difference $\Delta h$, between the free surfaces at the two branches of the tube. This plot is for $f=15\,Hz$ and $\Gamma=2.75$.}
\label{fig2}
\end{center}
\end{figure}

The excitation signal used was a sine wave.  The maximum amplitude of oscillation that could be reached was $(11.1 \pm 0.1)\,mm$. The excitation was controlled considering the maximum dimensionless acceleration defined as $\Gamma = A(2\pi f)^2/g$, where $g$ denotes the acceleration of gravity, $A$ the amplitude of the oscillations and $f$ the frequency of the oscillations. The value of $\Gamma$ was varied from $1$ to $5.75$. Lower values did not generate any instability and higher values of this quantity were not possible to achieve due to the limitations of the equipment.

\section{Results}

The height difference $\Delta h$ between the granular level in both branches of the U-tube grows exponentially with time (see Fig. \ref{fig2}). The growth rate $\gamma$ is determined from the fit of the exponential $\Delta h = \Delta h_0e^{(\gamma t)}$, where $\Delta h_0$ is the initial height difference and $t$ is the time.

\begin{figure*}
\begin{center}
\includegraphics[width=0.6\textwidth]{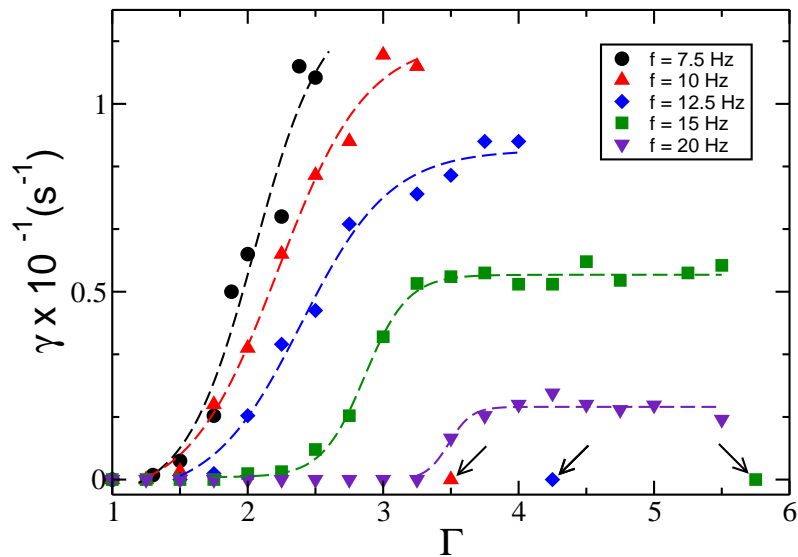}
\caption{Growth rate $\gamma$ versus the maximum adimensional acceleration $\Gamma$. The dashed lines correspond to sigmoidal fits of each curve given by equation (1). Each point represents the average of ten similar experiments, the error bars have been omitted to ease visualization. The three arrows in the graph are pointing to the values ($\Gamma, \gamma$) that mark an abrupt transition from nonzero $\gamma$ to zero growth rate at sufficiently large $\Gamma$.}\label{fig3}
\end{center}
\end{figure*}

The graph of the growth rate $\gamma$ versus dimensionless $\Gamma$ is shown in  Fig.\ref{fig3}. The horizontal axis begins at $\Gamma = 1$, because for $\Gamma < 1$, $\gamma = 0$. For each curve the oscillation frequency is fixed and the amplitude is varied. For all frequencies investigated there was a seemingly continuous transition from zero growth at the lowest amplitudes to a finite $\gamma$ at higher amplitudes. It is apparent that for the larger frequencies ($15\,Hz$ and $20\,Hz$) the growth rate reaches a saturation value, for sufficiently large amplitudes. For three of the intermediate frequencies ($10\,Hz$, $12.5\,Hz$ and $15\,Hz$) we see that $\gamma$ reaches a point where it changes abruptly to zero at a threshold of $\Gamma$ (see arrows in Fig. \ref{fig3}). This suggests the occurrence of a first order phase transition,  but more work needs to be done to characterize this observed abrupt transition. For the lowest frequency measured ($7.5\,Hz$) we could not determine whether there is a saturation region or a transition from high nonzero growth rate to zero growth at high amplitudes because of the limitations of our equipment.

Our experimental data is well described by a Boltzmann sigmoidal curve for $\gamma$ versus $\Gamma$. This fitting relation is given by eq. (\ref{eq1}), and it has four fitting parameters: (a) The higher asymptote $\gamma_{max}$ (characteristic value for the higher growth rates) (b) The size of the interval $\Delta\gamma$ (gives the range of growth rate from the lower asymptote to upper asymptote); (c) the inflection point $\Gamma_i$; and (d) the width $\Delta\Gamma$ of the transition region.

\begin{center}
\begin{equation}\label{eq1}
\gamma=\gamma_{max}-\left[\frac{\Delta\gamma}{1+e^{(\Gamma-\Gamma_i)/\Delta\Gamma}}\right].
\end{equation}	
\end{center}

\begin{figure}[!ht]
\begin{center}
\includegraphics{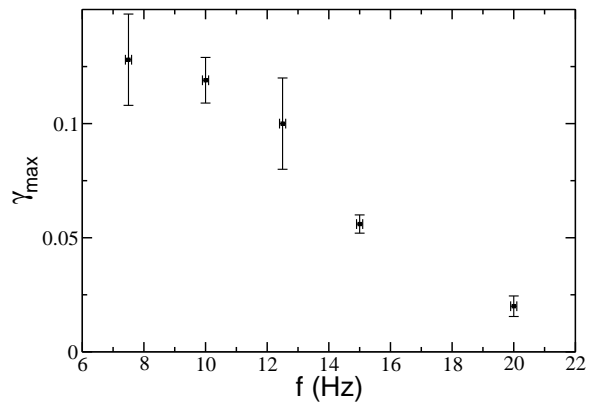}
\caption{Plot of the maximum growth rate $\gamma_{max}$ as a function of frequency. It is clear that higher frequencies reduce the ability of the granular material to flow upwards.}\label{fig4}
\end{center}
\end{figure}

A plot of $\gamma_{max}$ as a function of the frequency is shown in Fig. \ref{fig4}. This parameter is the saturation growth rate and gives an upper limit for the speed of the upward granular transport through the tube. We observe that the ability of the granular material to flow upwards is reduced as the frequency increases. For higher frequencies we can expect that the granular system approaches a liquid like behavior, and consequently, as the frequency rises it becomes more difficult for the granular bed to climb up.

For each frequency we can separate the observed behavior in Fig. \ref{fig3} in three regions. One for which $\gamma=0$ at low $\Gamma$, the second one for which $\gamma \neq 0$, for intermediate values of $\Gamma$; and a third region for which $\gamma = 0$ and $\Gamma$ is high. We can speculate that, for a given frequency, at low $\Gamma$; the granular bed behaves like a solid; at intermediate values of $\Gamma$; a mixed state occurs were the solid like state and the fluid state combine in a complex way to produce the collective upward motion of the grains in one of the branches of the container; and for sufficiently high $\Gamma$ the liquid like state dominates and the grains are unable to climb up on one of the branches of the tube.

We will now focus below on two types of transitions that were observed. One of them was a (seemingly continuous) transition from zero growth at the lowest amplitudes to a finite $\gamma$ at higher amplitudes. The other was a ``jump'' in the value of $\gamma$ (resembling a first order phase transition), which marked a transition from finite growth rate to no growth, when $\Gamma$ was sufficiently large. A detailed study of each transition requires further investigations. For the case of the abrupt transition, a different experimental setup capable of reaching larger amplitudes is necessary. Nevertheless, our data could give us an insight on the nature of the transition at low $\Gamma$.

\begin{figure}[!ht]
\begin{center}
\includegraphics{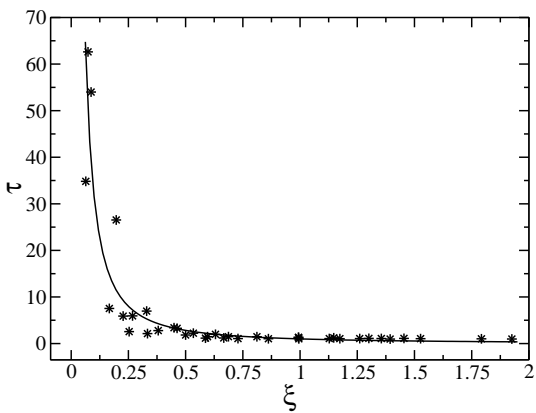}
\caption{The U-tube instability seems to exhibit a continuous phase transition. Here we can see how the characteristic time $\tau(\xi)$ diverges as $\Gamma$ approaches a critical value $\Gamma_c\approx 0$. The data shown correspond to all the non zero $\gamma$ measured.}
\label{fig5}
\end{center}
\end{figure}

To probe the nature of the transition at low $\Gamma$, we plot in Fig. \ref{fig5}, $\tau = \gamma_{max}/\gamma$ versus $\xi = (\Gamma-\Gamma_c)/\Gamma_c$. The parameter $\tau$ can be thought as a characteristic relaxation time for the evolution from a symmetric to an asymmetric configuration. The critical value $\Gamma_c$ of the adimensional acceleration was taken as the mean between the highest $\Gamma$ before $\gamma$ becomes non zero, and the lowest value of $\Gamma$ for which $\gamma$ is different from zero. At the lower amplitudes it was observed that the critical value $\Gamma_c$ for the transition from zero growth to nonzero $\gamma$, shifted to higher values of $\Gamma$ as the frequency increased. We can see how $\tau$ diverges as $\Gamma$ approaches the critical value $\Gamma_c$. More experiments have to be done before we can determine the nature of this transition.

Finally, we want to concentrate on the region of nonzero growth, for each curve we fixed the frequency and changed the amplitude $A$ of vibration, assuming that $\gamma_{min}\approx 0$, we can rewrite equation (\ref{eq1}) in the following form:

\begin{center}
\begin{equation}\label{eq2}
\gamma_n(\Gamma) = \frac{1}{1+e^{(\Gamma_i-\Gamma)/\Delta\Gamma}},
\end{equation}
\end{center}

\noindent where $\gamma_n = \gamma/\gamma_{max}$.

This expression gives the value of the normalized growth rate as a function of $\Gamma$. For each frequency equation (\ref{eq2}) is a solution of the following differential equation:

\begin{center}
\begin{equation}\label{eq3}
\frac{d\gamma_n}{d\Gamma}=\frac{1}{\Delta\Gamma}\left[1-\gamma_n\right]\gamma_n.
\end{equation}
\end{center}

\begin{figure}[!ht]
\begin{center}
\includegraphics{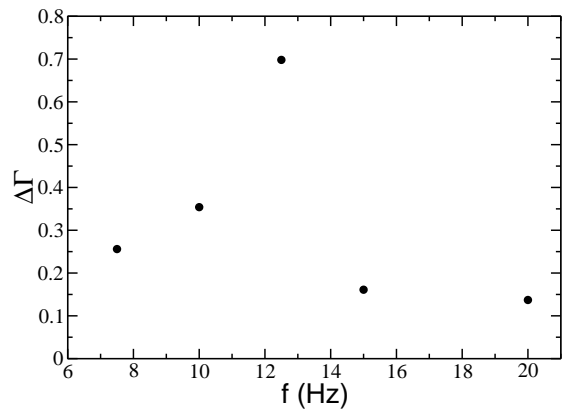}
\caption{Plot of $\Delta\Gamma$ versus the frequency of vibration. The parameter $\Delta\Gamma$ is a measure of the increase in energy supplied to the granular bed through vibrations, to go from zero growth to the plateau of the curve $\gamma(\Gamma)$ (see fits in Fig. \ref{fig3}) for a given frequency. A sharp increase occurs around $12.5\,Hz.$}\label{fig6}
\end{center}
\end{figure}

This equation is analogous to the well know logistic equation so we can picture $\gamma_n = 1$ as a stable fixed point for $\gamma_n (\Gamma)$. Therefore as $\Gamma$ increases the growth rate approaches its maximum value (which correspond to the stable fixed point). The smaller $1/\Delta\Gamma$, the more energy needs to be supplied through vibration to reach the plateau of the curve $\gamma_n(\Gamma)$. In Fig. \ref{fig6} we plotted the width $\Delta\Gamma$ of the region where $\gamma$ grows. We observe that $\Delta\Gamma$ rises with the frequency up to $f = 12.5\,Hz$ and then drops almost an order of magnitude. This behavior signals a qualitative change around  $f = 12.5\,Hz$. for this frequency $1/\Delta\Gamma$ is much smaller than for the other ones we measured, therefore around this particular frequency we need to supply more energy to the granular bed to make it go from zero growth rate to the fastest growth. 

A physical model that would explain our data and the above relationship for the growth rate and the adimensional acceleration for a given frequency is beyond the scope of this paper, but we consider that the simplicity of the results should help to suggest the relevant conditions that need to be satisfied by a reasonable model for the behavior of a vertically shaken granular bed in a partially filled U-tube.

\section{Conclusions}

We have made an experimental characterization of the rate of growth for the height difference $\Delta h$ between the free levels of the granular columns in a partially filled U-tube subject to vertical vibrations. We have observed a continuous transition from zero growth to non zero growth rate $\gamma$. The region of finite growth was well described by a differential equation analogous to the logistic equation. We also observed an abrupt transition from high values of $\gamma$ to zero growth similar to a first order phase transition. 

\begin{acknowledgements}
This work has been done under the LOCTI project: ``La ac\'{u}stica como punta de prueba de los medios porosos y sistemas granulares'' and IVIC project 857. It was also supported in part by FONACIT under Grant S1-2000000624, and by DID of the Universidad Sim\'{o}n Bol\'{\i}var.
\end{acknowledgements}


\end{document}